# Quantifying Subtle Regions of Order and Disorder in Tumor Architecture by Calculating the Nearest-Neighbor Angular Profile

David H. Nguyen, Ph.D.
Affiliate Scientist
Dept. of Cellular & Tissue Imaging
Division of Molecular Biophysics and Integrated Bioimaging
Lawrence Berkeley National Laboratory
DHNguyen@lbl.gov

April 2017

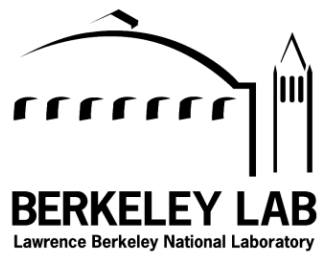

## Abstract

Pathologists routinely classify breast tumors according to recurring patterns of nuclear grades, cytoplasmic coloration, and large-scale morphological formations (i.e. streams of spindle cells, adenoid islands, etc.). The fact that there are large-scale morphological formations suggest that tumor cells still possess the genetic programming to arrange themselves in orderly patterns.  However, small regions of order or subtle patterns of order are invisible to the human eye. The ability to detect subtle regions of order and correlate them with clinical outcome and resistance to treatment can enhance diagnostic efficacy.  By measuring the acute angle that results when the line extending from the longest length within a nucleus intersects with the corresponding line of an adjacent nucleus, the degree of alignment between two adjacent nuclei can be measured. Through a series of systematic transformations, subtle regions of order and disorder within a tumor image can be quantified and visualized in the form of a heat map. This numerical transformation of spatial relationships between nuclei within tumors allows for the detection of subtly ordered regions.

## Graphical Abstract

**A** Longest Length Axes of Neighboring Nuclei Form Acute Angles

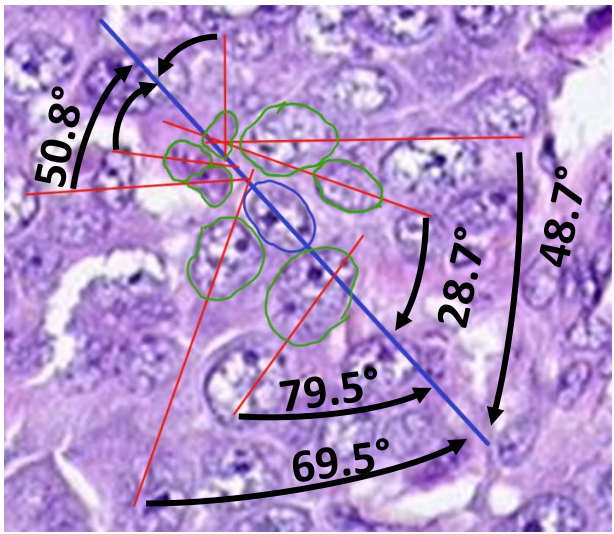

**B** Nearest-Neighbor Angular Profile (N-NAP) of a Single Nucleus

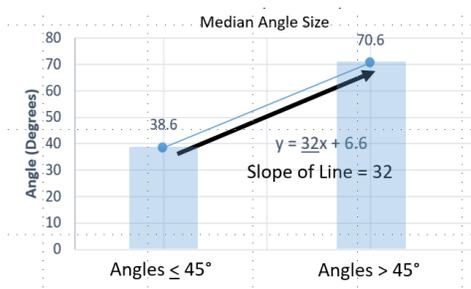

**C** Schematic of Concept

Slope of the N-NAP of Each Nuclei Represents Local Regions of Order or Disorder

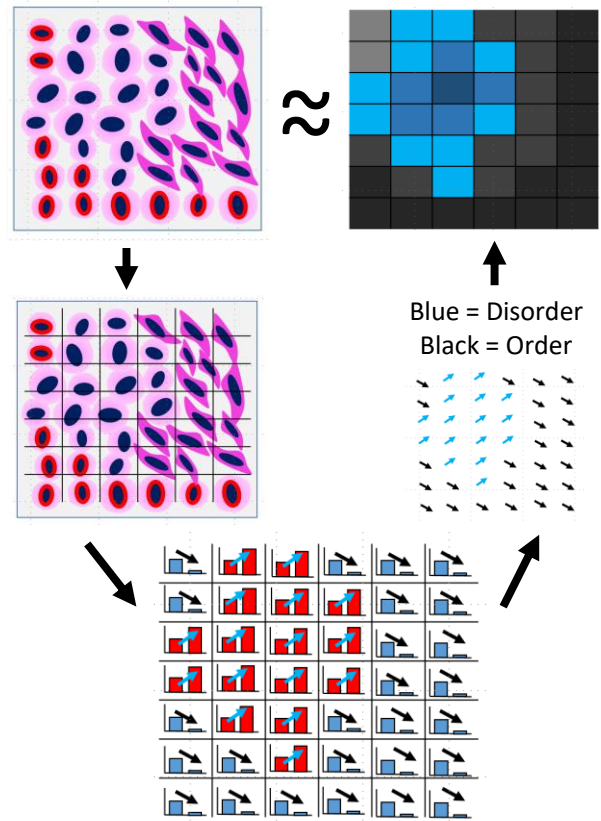

Blue = Disorder
Black = Order

# Introduction

**Order and Disorder in Tumor Architecture**

The fact that pathologists can classify neoplastic lesions by recurring large-scale morphological features suggests that there is a certain degree of order within the architecture of a tumor [1]. However, beyond broad classifications such as adenocarcinoma, apocrine carcinoma, metaplastic carcinoma, medullary carcinoma, and other categories (Figure 1), the human eye cannot detect smaller regions of order within a tumor slice visualized by hematoxylin & eosin (H&E) stains [2]. Subtle changes in architecture are hard with measure. Without numerical analysis, statistical approaches cannot quantify subtle changes in local regions within a tumor.

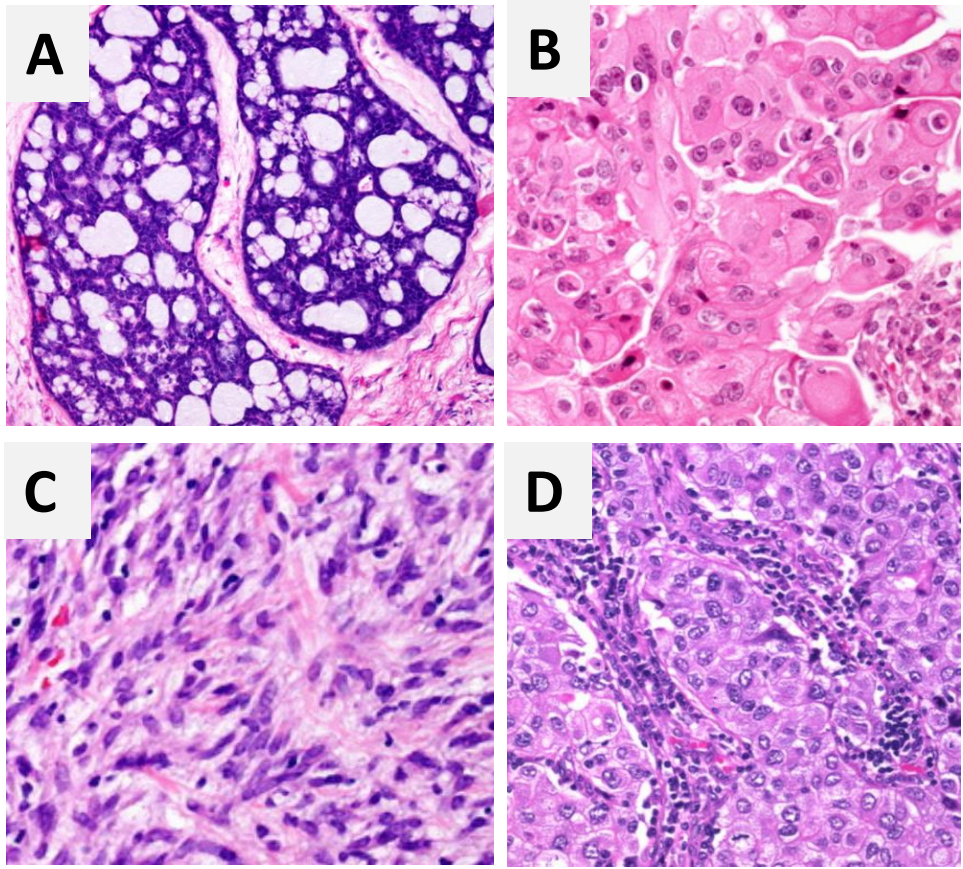

**Figure 1. Examples of Histopathological Classifications of Breast Tumors.** The magnification of each image is varied in order to highlight distinct nuclear and cellular features. Tumors were visualized by hematoxylin & eosin stains. (A) Adenoid cystic carcinoma. (B) Apocrine carcinoma. (C) Metaplastic carcinoma (spindle cell carcinoma). (D) Medullary carcinoma. Images adapted from WebPathology.com.

# Nearest-Neighbor Angular Profiling (N-NAP) Detects Subtle Changes in Tissue Architecture

Normal mammary epithelia have their nuclei aligned side-by-side, with the longest length nearly parallel with each other (Figure 2). However, since neighboring nuclei are not uniformly aligned, the lines that extend from their longest length intersect each other at acute angles. By measuring the angles formed by the lines that extend from the maximum nuclear axis of two adjacent nuclei, the visual near-uniformity can be represented numerically in the form of degrees. This quantitative approach allows for detection of subtle changes in nuclear packing that are invisible to the human eye. This method will be referred to as the Nearest-Neighbor Angular Profile (N-NAP).

Breast tumors are a mixture of multiple cell types, including ductal epithelia, fibroblasts, vasculature, nerves, and immune cells [3]. Though the nuclear morphology of epithelial cells in breast cancers are distinct from other cells types and can easily be distinguished by a pathologist, the arrangement of cells and their nuclei can appear random or highly complex, depending on other visual features present in normal mammary ducts.

Though arranged in ways that do not intuitively suggest a pattern, the alignment of adjacent nuclei in tumors can be quantified using the N-NAP approach. Pathologists routinely identify large morphological features in tumors (Figure 1), such papillary (finger-like) projections, adenoid (circular clusters) foci, and cells arranged as if flowing in a fluid path (referred to as the "streaming" effect) [2]. These large pathological features are the result of tumor cells attempting to do what normal epithelial cells in the mammary gland do. They are trying to form hollow, branching ducts [4]. Tumor cells that are not forming the large pathological features are assumed to be in a random or non-ordered pattern.

However, if breast tumor cells still possess the programming to be like normal breast epithelial cells, though the programming is not fully functional, it can be hypothesized that the tumor cells are attempting to order themselves on a micro scale. This scale would be too small to be detected by the human eye as more than random arrangements. However, a precise quantitative approach applied to a large sample size would be able to detect subtle regions of order amid a sea of seeming randomness. Quantitative approaches have successfully detected morphological subtly in nuclei and tissues [5, 6].

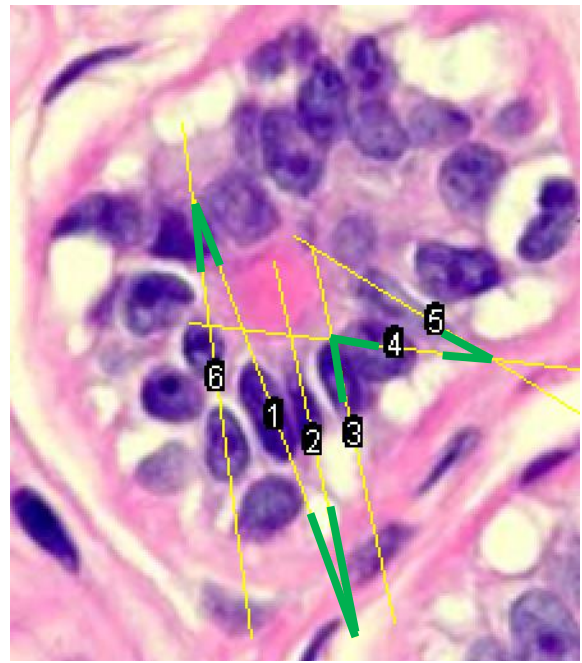

**Figure 2. Nuclei of normal breast epithelia are aligned nearly parallel next to each other.** Normal human breast lobule (adapted from PathPedia.com). Nuclei that are adjacent to each other and aligned in the same direction have longest length axes that intersect to form acute angles. The closer to having parallel alignment, the smaller the angle formed by the longest length axes will be (green angles). The misalignment of cells resulting from ductal hyperplasia will cause the average angle size to be larger than that found in normal ducts. The nearest-neighbor alignment profile is useful for detecting subtle changes in alignment.

In order to be systematic, a nucleus' nearest neighbors are defined as those within a radius extending from the edge of the nucleus. The nucleus of interest is referred to as the "central nucleus," while the nuclei around it are referred to as neighbors. To objectively define a ring for detecting the nearest neighbors around a central nucleus, the ring's width is defined to be the longest length within the central nucleus (Figure 3A,C).

Since there may be fewer neighboring nuclei around some central nuclei due to the total size of the cells in that location, a second ring was defined in order to include more neighboring nuclei. The width of the second ring, which is concentric to the first ring, is also defined as the longest length within the central nucleus (Figure 3B,D).

# Primary Neighbors | Secondary Neighbors

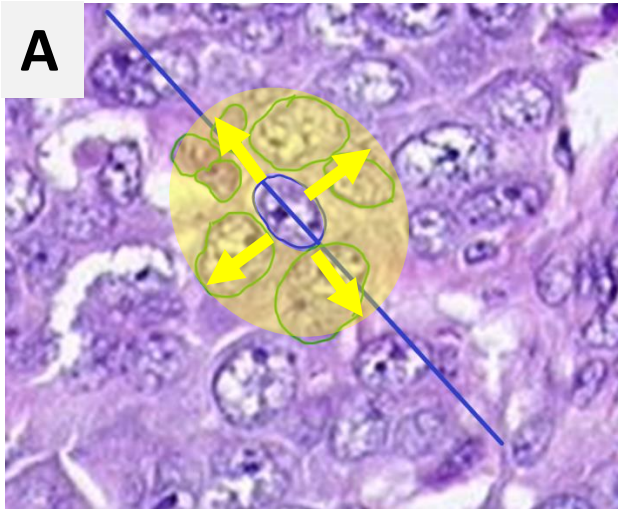
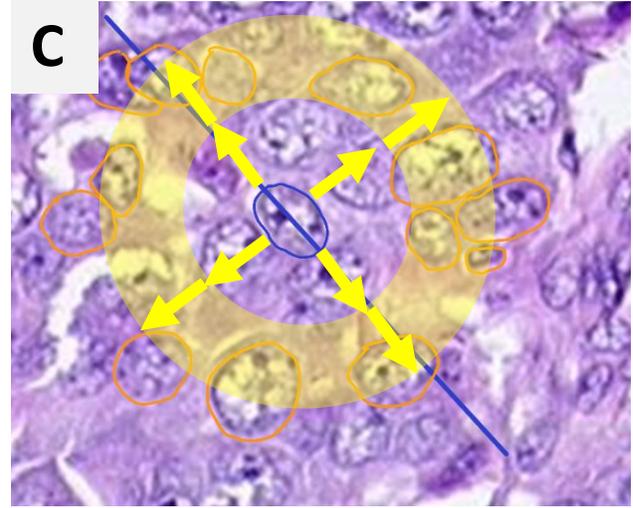
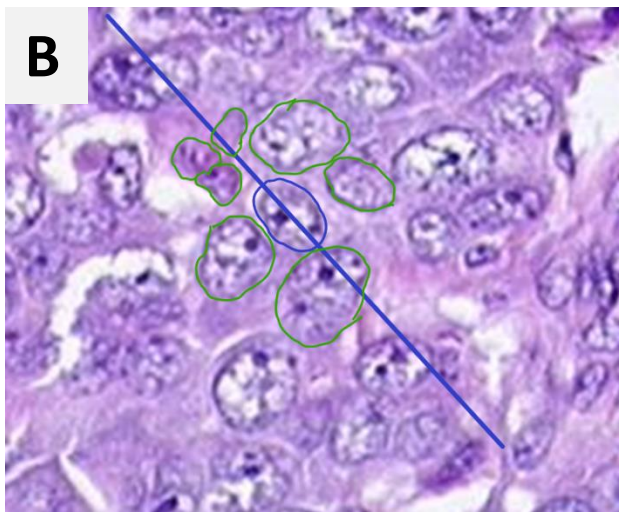
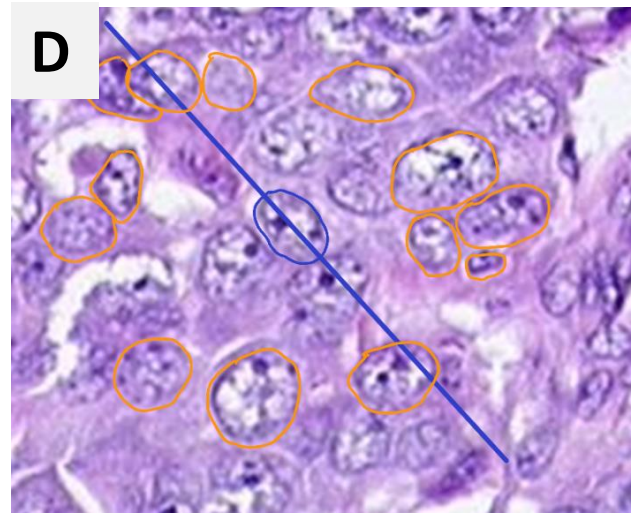

**Figure 3. Defining the nearest neighbor nuclei around a central nuclei.**
Central nuclei are outlined in blue. Their longest length axis is also in blue. Neighboring nuclei in the primary ring are outlined in green. Neighboring nuclei in the secondary ring are outlined in orange. (A) The length of yellow arrows is defined as the longest length within the central nuclei. The yellow ring highlights the region where primary nuclei are located. (B) As in A, but without yellow highlighting. (C) The yellow ring defines the region where secondary neighboring nuclei are located. The width of the second ring is also defined as the length of the longest line within the central nucleus. (D) As in C, but without yellow highlighting.

| Primary Neighbors | Secondary Neighbors |
|---|---|
| 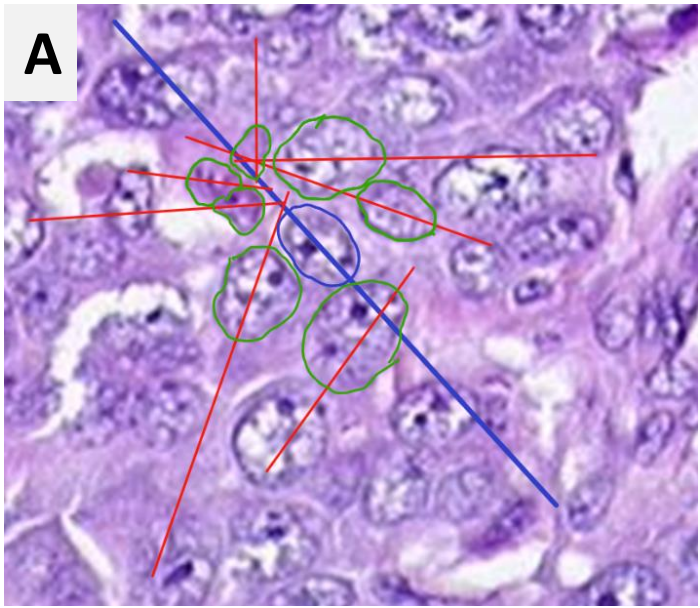 A | 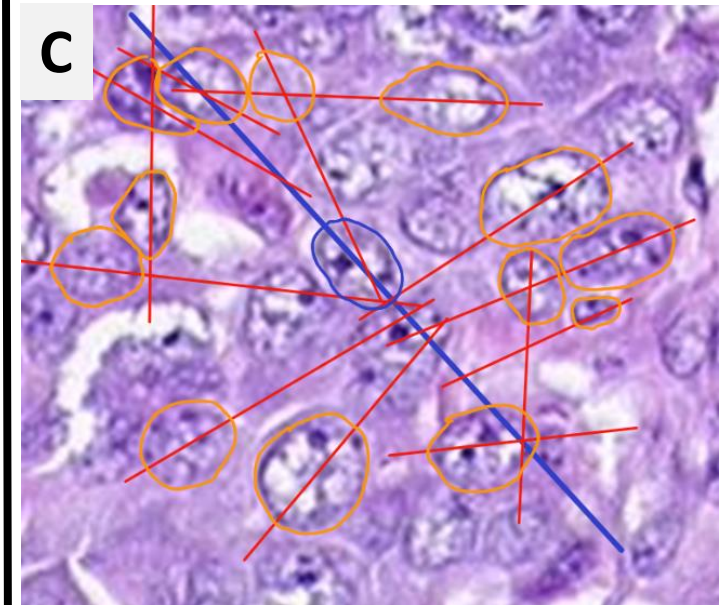 C |
| 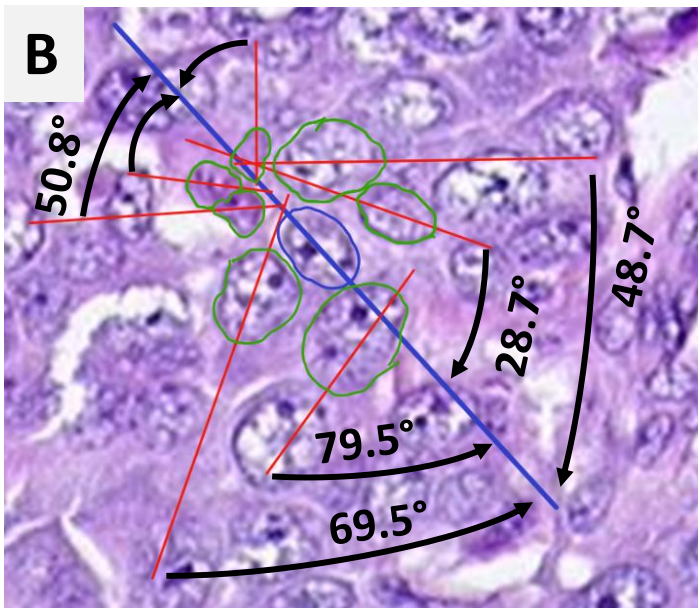 B | 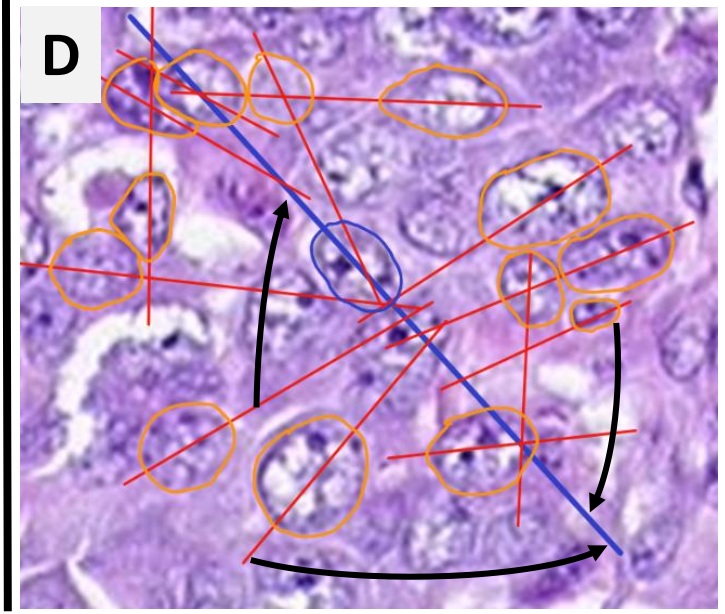 D |

**Figure 4. The arrangement of neighboring nuclei around a central nuclei can be quantified by measuring the acute angle resulting from the intersection of the lines extending from each nucleus' longest length axis.** (A) The central nucleus is outlined in blue and it's the line extending from its longest axis is also in blue. The nearest neighbor nuclei in the primary ring are outlined in green. Red lines represent a extensions of the axes of the longest lengths within each neighbor nuclei. (B) The acute angle resulting from the intersection of each red line to the blue line of the central nucleus is measured. The angle is a representation of the degree of alignment between two adjacent nuclei. (C) Secondary neighbors are outlined in orange. (D) As in B, the acute angle resulting from the intersection of each red line with the blue line is measured.

**Nearest-Neighbor Angular Profile (N-NAP)**

Plotting the measure of the angles between a central nucleus and its primary and secondary nearest neighbors creates a numerical representation of the spatial relationships between nuclei that cluster near each other. (Figure 4B, D). Since the nuclei of luminal epithelial cells are packed nearly parallel with each other (Figure 5C), the angle between adjacent nuclei will tend to be <45°.

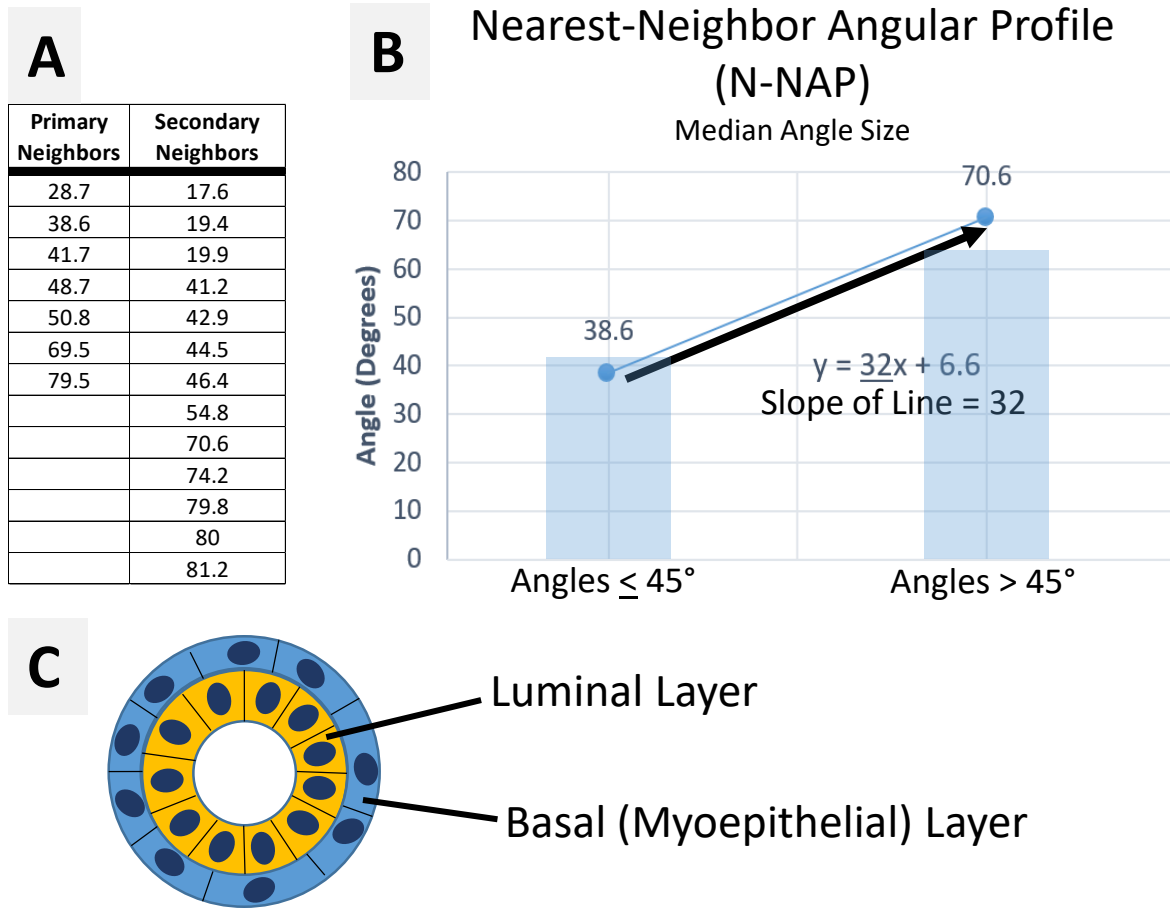

**Figure 5. The arrangement of two adjacent nuclei can be quantified by the acute angle that results when the lines from their longest axes intersect.** (A) The measures of each angle between nearest-neighbor nuclei and the central nucleus, in Figure 4B and 4D, respectively. The data was ranked from smallest to largest. (B) The angles of both primary and secondary nearest-neighbors were binned into those that were ≤45° or that were >45°. This was based on the assumption that most angles between adjacent nuclei in the luminal compartment of normal mammary ducts is often <45°. This assumption is subject to change depending the prior knowledge of how the normal tissue, from which the cancer is derived, is organized. The slope of the line between the two median values can be used as measure of the "degree" of disorder surrounding the central nucleus. This allows for an image of a tumor to be transformed into a heat map based on the magnitude of the slope surrounding each nuclei. (C) Schematic of the cross section of a mammary duct showing the two compartments of the epithelial wall. Luminal cells (yellow) are packed side-by-side like bricks. Their nuclei are similarly aligned. Myoepithellial cells form the basal layer (blue) and exhibit a flatter, stretched shape.

## In Principle: N-NAP Can Detect Local Regions of Order and Disorder in a Tumor

Tumors exhibit complex cellular and stromal architecture. Figure 6 shows a schematic of the types of cellular and nuclear patterns that are commonly observed in H&E sections of breast tumors. The most obvious regions of orderly arrangement are the streaming effect of elongated spindle cells (dark pink) and cells that are aligned side-by-side as if forming a wall (red bordered nuclei). However, the complex architecture inside of a tumor can make these small regions of orderliness seem random.

This paper proposes that the N-NAP method of quantifying the alignment of neighboring nuclei can reveal subtle regions of orderliness throughout the image of a tumor. By converting recurring visual patterns systematically into numerical relationships, subtle regions of order can be assessed by statistical methods.

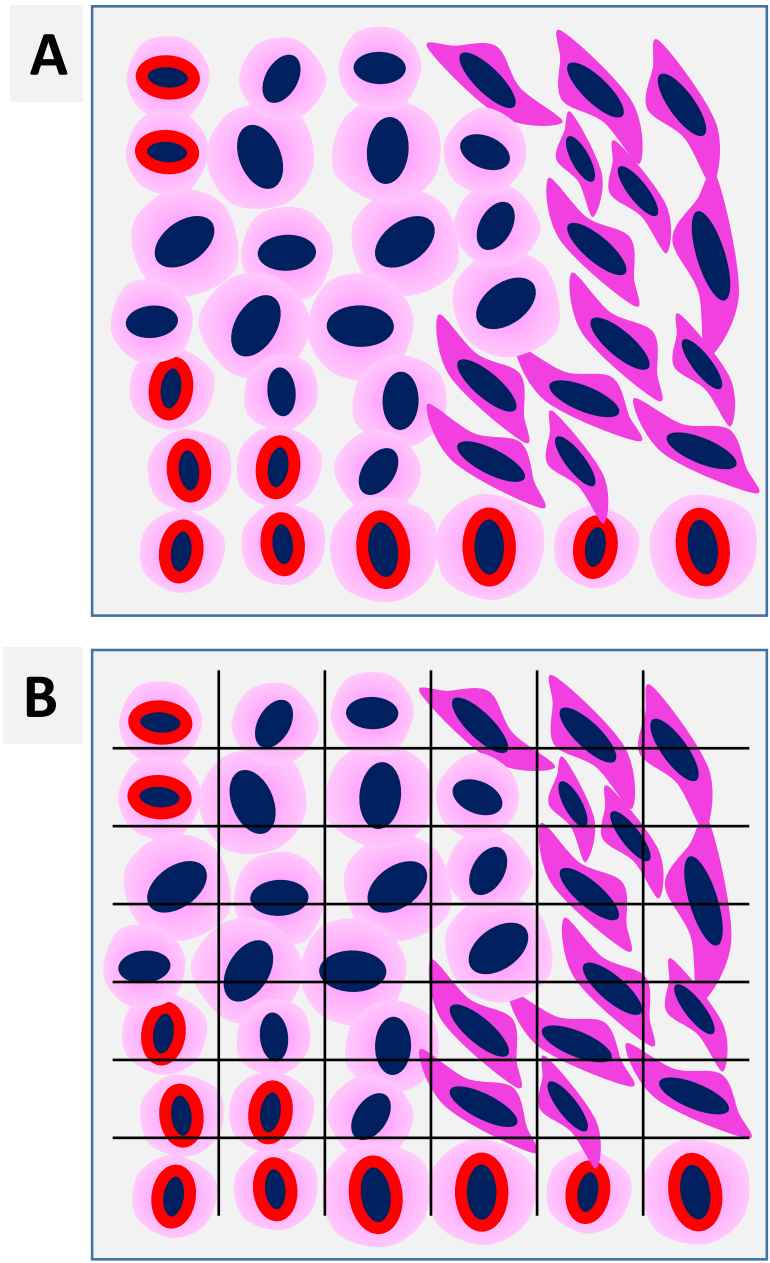

**Figure 6. Schematic of regions of order and disorder in a tumor.** (A) Two types of ordered arrangement are streams (dark pink, spindle cells) and side-by-side arrangements (cells that have nuclei with red borders). (B) Quantifying the N-NAP of each nuclei in an tumor is like placing a grid of the tumor that represents pixels in an image.

## Spatial Annotation of N-NAPs in a Tumor Can Reveal Local Regions of Order or Disorder

N-NAPs contain two categories of angles, those less than or equal to 45° and those that are greater than 45°. The rationale behind the 45° threshold is that luminal epithelial cells in normal breast ducts are aligned such that the N-NAP angles between adjacent nuclei are often <45°.

This threshold results in cells that have many misaligned neighbors to exhibit N-NAP profiles that have more angles that are >45°. Thus, the slope of the line connecting the two bins can treated as a transformation of the N-NAP profile data.

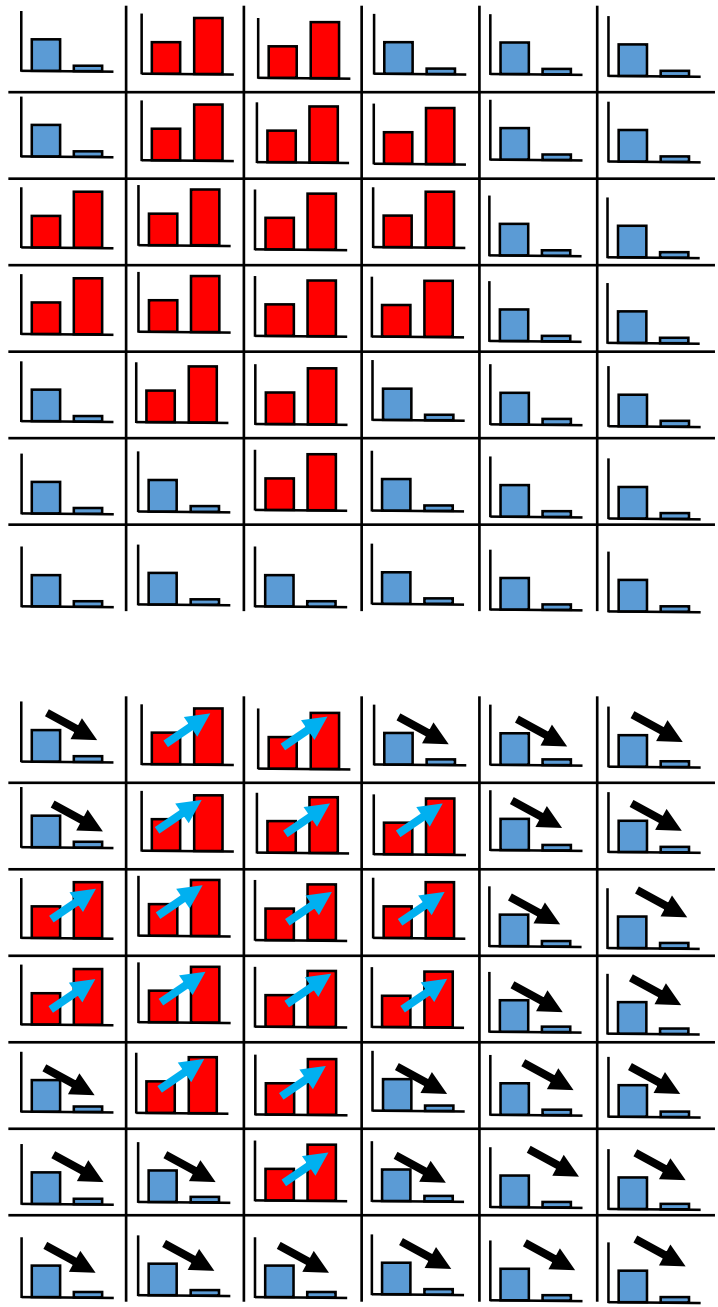

Figure 7. Schematic of how N-NAP profiles can be a representation of the tumor image.

## The Vector Space Derived From Spatially Depicted N-NAPs Can be Converted to 2D Heat Maps Representing Regions of Order or Disorder

The blue vectors represent regions of misaligned nuclei, which are regions wherein neighboring nuclei exhibit a disorderly arrangement, as compared to the nearly parallel packing of normal breast epithelial cells.

The black vectors represent regions wherein neighboring nuclei are aligned in a ordered spatial pattern. The black vectors do not specify what type of pattern is present, just that there is a orderliness to nuclear alignment in that area.

Plotting the magnitude of the slope of each vector can reveal the degree of intensity that adds depth of the heat map (Figure 9).

This method allows pathologists to quantify an index that reveals the areas of subtle order that was previously hidden to the human eye. This method also allows previously hidden pathological features to be dissected by numerous mathematical approaches. Statistical analysis of tumor architecture can potentially enhance personalized medicine.

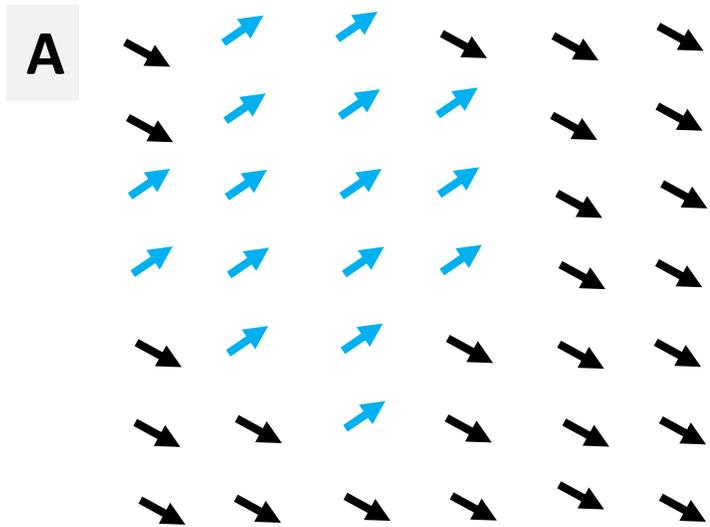
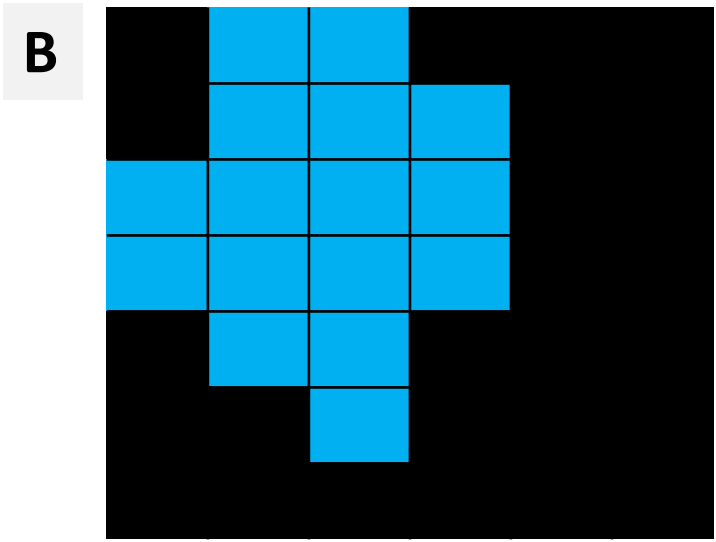
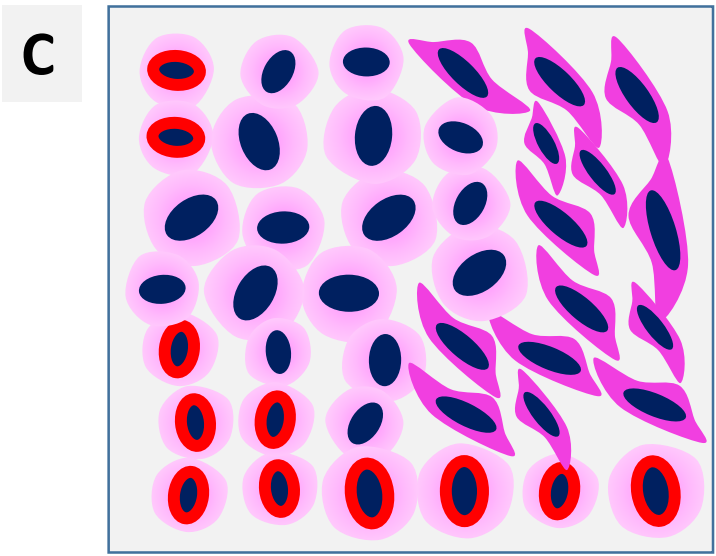

**Figure 8. Schematic of how N-NAP profiles can represent a tumor image in the form of vectors and squares.** (A) Overlaying the tumor image with the lines connecting the two bins in each N-NAP profile creates a vector map that represents the tumor. (B) Converting each vector in to a square area of arbitrary units creates a heat map that represents the original image. (C) The original image of the hypothetical tumor, for comparison with the hypothetical heat map in B.

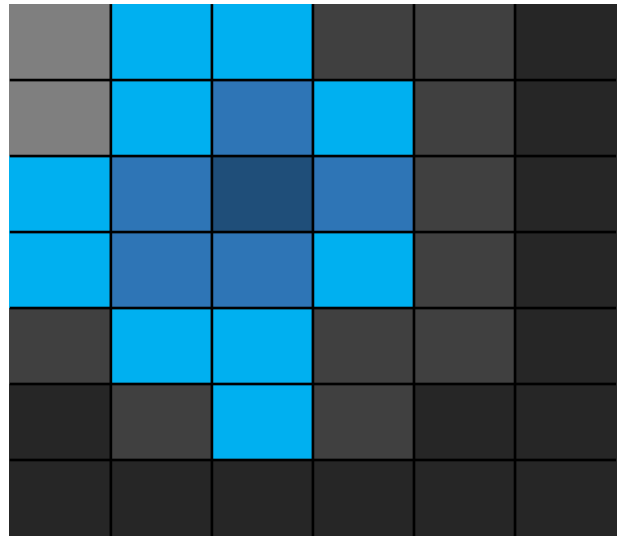

**Figure 9. The vector space derived from spatially depicted N-NAPs can be converted to 2D heat maps or 3D surface plots.** Hypothetical heat map based on varying slopes of N-NAP profiles.